\documentclass{article} 
\usepackage{nips12submit_e,times}
\usepackage{multirow}
\usepackage{times}
\usepackage{amsmath,amssymb}
\usepackage[amsmath,thmmarks]{ntheorem}
\usepackage{graphicx}
\usepackage{listings}             
\usepackage{color}
\usepackage{cite}
\newtheorem{thm}{Theorem}

\newtheorem{cond}{Condition}

\newtheorem{lem}{Lemma}
\newtheorem{exa}{Example}

\newtheorem{crol}{Corollary}

\newcommand{\Omit}[1]{}
\newcommand{\lirong}[1]{\footnote{\color{red}Lirong: #1}}

\newenvironment{proof}{\vspace{-2mm}{\bf Proof:}}{\hfill $\Box$ }
\newenvironment{sketch}{\vspace{-2mm}{\bf Proof sketch:}\rm }{\hfill $\Box$ }

\newcommand{\mc}{\mathcal C}

\newcommand{\pr}{\text{Pr}}

\nipsfinalcopy
\begin{document}
\title{Random Utility Theory for Social Choice}
\author{Hossein Azari Soufiani\\ SEAS, Harvard University\\ azari@fas.harvard.edu \And
David C.~Parkes \\ SEAS, Harvard University\\parkes@eecs.harvard.edu \And
Lirong Xia\\ SEAS, Harvard University\\lxia@seas.harvard.edu
}

\maketitle

\begin{abstract}
{\em Random utility theory} models an agent's preferences on
alternatives by drawing a real-valued score on each alternative
(typically independently) from a parameterized distribution, and then
ranking the alternatives according to scores. A special case that has received
significant attention is the Plackett-Luce model, for which fast
inference methods for maximum likelihood estimators are
available. This paper develops conditions on general
random utility models that enable fast inference within
a Bayesian framework through MC-EM, providing concave log-likelihood
functions and bounded sets of global maxima solutions. Results on both real-world and simulated data provide support
for the scalability of the approach and capability for model selection among general random utility models including Plackett-Luce.
\end{abstract}

\section{Introduction}

Problems of learning with rank-based error
metrics~\cite{Liu11:Learning} and the adoption of learning for the
purpose of rank aggregation in social
choice~\cite{Conitzer05:Common,Conitzer09:Preference,Xia10:Aggregating,Xia11:Maximum,Roos11:How,Procaccia12:Maximum}
are gaining in prominence in recent years.  In part, this is due to
the explosion of socio-economic platforms, where opinions of users
need to be aggregated; e.g., judges in crowd-sourcing contests, ranking
of movies or user-generated content.

In the problem of social choice, users submit ordinal preferences
consisting of partial or total ranks on the alternatives and a single
rank order must be selected to be representative of the reports.
Since
Condorcet~\cite{Condorcet1785:Essai}, one approach to this problem is
to formulate social choice as the problem of estimating a true
underlying world state (e.g., a true quality ranking of alternatives),
where the individual reports are viewed as noisy data in regard to the
true state. In this way, social choice can be framed as a problem of
inference.

In particular, Condorcet assumed the existence of
a true {\em ranking} over alternatives,
with a voter's preference between any pair of alternatives
$a, b$
 generated to agree with the true ranking
with probability $p>1/2$ and disagree otherwise.
Condorcet proposed to choose as the outcome of social
choice the
ranking that maximizes the likelihood of observing the voters'
preferences. Later, Kemeny's rule was shown to provide
the maximum likelihood
estimator (MLE) for this model~\cite{Young95:Optimal}.

But Condorcet's probabilistic model
assumes identical and independent distributions on pairwise comparisons.
This ignores the strength in agents' preferences (the same probability $p$
is adopted for all pairwise comparisons), and allows for cyclic preferences.
In addition, computing the winner through the Kemeny
rule is
$\Theta_2^P$-complete~\cite{Hemaspaandra05:Complexity}.

To overcome
the first criticism, a more recent literature
adopts the {\em random utility model} (RUM) from
economics~\cite{Thurstone27:Law}. Consider $\mc=\{c_1,..,c_m\}$
alternatives.
In RUM, there is a ground truth utility (or score) associated with
each alternative. These are real-valued parameters, denoted by
$\vec \theta=(\theta_1,\ldots,\theta_m)$.  Given this, an
agent
independently samples a random utility ($X_j$) for each alternative
$c_j$ with conditional distribution $\mu_j(\cdot|\theta_j)$.

Usually $\theta_j$ is the mean of $\mu_j(\cdot|\theta_j)$.\footnote{$\mu_j(\cdot|\theta_j)$ might be parameterized by other parameters, for example variance.} Let $\pi$
denote a permutation of $\{1,\ldots,m\}$, which naturally corresponds
to a linear order: $[c_{\pi(1)}\succ c_{\pi(2)}\succ\cdots\succ
  c_{\pi(m)}]$. Slightly abusing notation, we also use $\pi$ to denote
this linear order.
Random utility $(X_1,\ldots,X_m)$ generates a
distribution on preference orders, as
\begin{eqnarray}\label{Order1}
\pr(\pi\ |\ \vec{\theta})=\pr(X_{\pi(1)}>X_{\pi(2)}>\ldots>X_{\pi(m)})
\end{eqnarray}

\begin{figure*}[h]
  \includegraphics[width=1\textwidth]{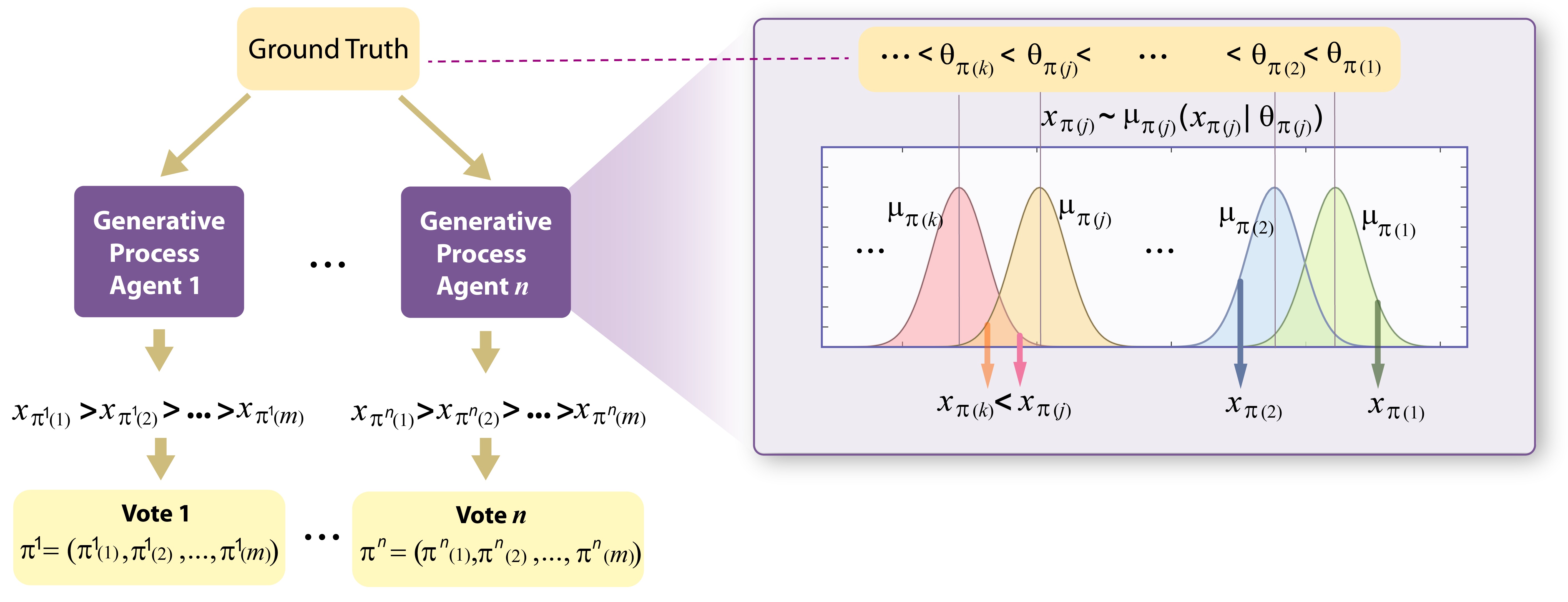}\vspace{-5mm}
  \caption{\small The generative process for RUMs. \label{fig:rum}}
\end{figure*}

The generative process is illustrated in Figure~\ref{fig:rum}.

%

Adopting RUMs rules out cyclic preferences, because each agent's
outcome corresponds to an order on real numbers, and it also
captures the strength of preference, and thus overcomes the second
criticism, by assigning a different parameter ($\theta_j$) to each
alternative.

A popular RUM  is Plackett-Luce (P-L)~\cite{Luce59:Individual,Plackett75:Analysis}, where the random utility terms are generated according to Gumbel distributions with fixed shape parameter~\cite{Block60:Random,Yellott77:Relationship}.
For P-L, the likelihood function has a simple analytical solution, making MLE inference tractable. P-L has been extensively applied in econometrics~\cite{McFadden74:Conditional,Berry95:Automobile}, and more recently in machine learning and information retrieval
(see~\cite{Liu11:Learning} for an overview). Efficient
methods of EM inference~\cite{Hunter04:MM,Caron12:Efficient}, and more recently expectation
propagation~\cite{Guiver09:Bayesian}, have been developed for P-L and its variants.

In application to social choice, the P-L model
has been used to analyze political elections~\cite{Gormley09:Grade}. EM algorithm has also been used to learn the {\em Mallows} model, which is closely related to the Condorcet's probabilistic model~\cite{Lu11:Learning}.

Although P-L overcomes the two difficulties of the Condorcet-Kemeny approach,
it is still quite restricted,
by assuming that the random utility terms are distributed as Gumbel, with each alternative is characterized by one parameter, which is the mean of its corresponding distribution.
In fact, little is known about inference in RUMs beyond P-L. Specifically, we are not aware of either an analytical solution or an efficient algorithm for MLE inference for one of the most natural models proposed by Thurstone~\cite{Thurstone27:Law}, where each $X_j$ is normally distributed.

\subsection{Our Contributions}
\vspace{-2mm}
In this paper we focus on RUMs in which the random utilities are independently generated with respect to distributions in the {\em  exponential family} (EF)~\cite{Morris12:Natural}. This extends the P-L model, since the Gumbel distribution with fixed shape parameters belonging to the EF.
Our main theoretical contributions are Theorem~\ref{thm1} and Theorem~\ref{thm2}, which propose conditions such that the log-likelihood function is concave and the set of global maxima solutions is bounded for the {\em location family}, which are RUMs where the shape of each
distribution $\mu_j$ is fixed and the only latent variables are the locations, i.e., the means of $\mu_j$'s. These results hold for existing special cases,
such as the P-L model, and many other  RUMs, for example the ones where each $\mu_j$ is chosen from Normal, Gumbel, Laplace and Cauchy.

We also propose a novel application of MC-EM.
We treat the
random utilities ($\vec{X}$) as latent variables,
and adopt the Expectation Maximization (EM)
method to estimate parameters $\vec{\theta}$.  The E-step for
this problem is not analytically tractable, and for
this we adopt a Monte Carlo approximation. We establish
through experiments that the Monte-Carlo error in
the E-step is
controllable and does not affect inference, as long as numerical
parameterizations are chosen carefully.
In addition, for the E-step we suggest a parallelization over the agents and alternatives and a Rao-Blackwellized method, which further increases the scalability of our method.

We generally assume that the data provides total orders on alternatives from
voters, but comment on how to extend the method
and theory to the
case where the input preferences are {\em partial} orders.

We evaluate our approach on synthetic data as well as two real-world datasets,
a public election dataset and one involving rank preferences
on sushi. The experimental results suggest that the approach is
scalable despite providing significantly improved modeling flexibility
over existing approaches.

For the two real-world datasets we have studied, we compare RUMs with normal distributions and P-L in terms of four criteria: log-likelihood,  predictive log-likelihood, Akaike information criterion (AIC), and Bayesian information criterion (BIC). We observe that when the amount of data is not too small, RUMs with normal distributions fit better than P-L.
Specifically, for the log-likelihood, predictive log-likelihood, and AIC criteria, RUMs with normal distributions outperform P-L with 95\% confidence in both datasets.

\vspace{-2mm}
\section{RUMs and Exponential Families}\label{rumef}
\vspace{-2mm}
In social choice, each agent $i\in \{1,\ldots,n\}$ has a strict preference order on alternatives. This provides the data for an inferential approach
to social choice. In particular. let $L(\mc)$ denote the set of all linear
orders on $\mc$.  Then, a {\em preference-profile}, $D$, is a set of $n$
preference orders, one from each agent, so that $D \in L(\mc)^n$.

A {\em voting rule} $r$ is a mapping that assigns to each preference-profile a set of
winning rankings, $r: L(\mc)^n\mapsto (2^{L(\mc)}\setminus\emptyset)$. In particular, in the case of ties the set of winning rankings may include more than a singleton
ranking. In the maximum likelihood (MLE) approach to social choice,
the preference profile is viewed as {\em data}, $D=\{\pi^1,\ldots,\pi^n\}$.

Given this, the probability (likelihood) of the data given ground truth $\vec\theta$ (and for a particular $\vec\mu$) is
$\Pr(D\ |\ \vec\theta)=\prod^n_{i=1} \Pr(\pi^i\ |\ \vec\theta),$
where,
\begin{align}\label{int}
\!\!P(\pi|\vec{\theta})\!=\!\!\!\int^{\infty}_{x_{\pi(n)}=-\infty}\int^{\infty}_{x_{\pi(n-1)}=x_{\pi(n)}}\!\!\!..\int^{\infty}_{x_{\pi(1)}=x_{\pi(2)}} \!\!\!\!\!\!\!\!\mu_{\pi(n)}(x_{\pi(n)})..\mu_{\pi(1)}(x_{\pi(1)}) dx_{\pi(1)}dx_{\pi(2)}..dx_{\pi(n)}
\end{align}

The MLE approach to social choice selects as the winning ranking that
which corresponds to the $\vec\theta$ that maximizes $\Pr(D\ |\
\vec\theta)$. In the case of multiple parameters that maximize the
likelihood then the MLE approach returns a set of rankings, one
ranking corresponding to each parameterization.

In this paper, we focus on probabilistic models where each $\mu_j$
belongs to the {\em exponential family (EF)}. The density
function for each $\mu$ in EF has the following format:
\begin{align}\label{expf}
\pr(X= x)&=\mu(x)=e^{\eta(\theta) T(x)-A(\theta)+B(x)},
\end{align}
where $\eta(\cdot)$ and $A(\cdot)$ are functions of $\theta$, $B(\cdot)$ is a function of $x$, and $T(x)$ denotes the sufficient statistics for $x$, which could be multidimensional.

\begin{exa}[Plackett-Luce as an RUM~\cite{Block60:Random}]
In the RUM, let $\mu_j$'s be Gumbel distributions. That is, for alternative $j\in \{1,\ldots,m\}$ we have $\mu_j(x_j|\theta_j)=e^{-(x_j-\theta_j)}e^{-e^{-(x_j-\theta_j)}}$. Then, we have:
%
$\pr(\pi\ |\ \vec \lambda)=\pr(x_{\pi(1)}> x_{\pi(2)}>..> x_{\pi(m)})\notag= \prod^m_{j=1}{\lambda_{\pi(j)}\over\sum^m_{j'=j} \lambda_{\pi(j')}}$
%
, where $\eta({\theta_j})=\lambda_j=e^{\theta_j}$, $T(x_j)=-e^{-x_j}$, $B(x_j)=-x_j$ and $A(\theta_j)=-\theta_j$.
This gives us the Plackett-Luce model.

\end{exa}

\vspace{-2mm}
\section{Global Optimality and Log-Concavity}\vspace{-2mm}
In this section,
we provide a condition on distributions that guarantees that the likelihood function~\eqref{int} is log-concave in  parameters $\vec \theta$. We also provide a condition under which the set of MLE solutions is bounded when any one latent parameter is fixed. Together, this guarantees the convergence of our MC-EM approach to a global mode with an accurate enough E-step. We focus on the  {\em location family}, which is a subset of RUMs where  the shapes of all $\mu_j$'s are fixed, and the only parameters are the means of the distributions. For the location family, we can write $X_j=\theta_j+\zeta_j$, where $X_j\sim\mu_j(\cdot|\theta_j)$ and $\zeta_j=X_j-\theta_j$ is a random variable whose mean is $0$ and models an agent's {\em subjective noise}. The random variables $\zeta_j$'s do not need to be identically distributed for all alternatives $j$; e.g., they can be normal with different fixed variances.

We focus on computing solutions ($\vec \theta$) to maximize the log-likelihood function,
\begin{align}
l(\vec{\theta};D)=\sum^n_{i=1} \log \Pr(\pi^i\ |\ {\vec\theta})
\end{align}\vspace{-2mm}
%
\begin{thm}\label{thm1}
For the location family, if for every $j\leq m$ the probability density function for $\zeta_j$ is log-concave, then $l(\vec{\theta};D)$ is concave.
\end{thm}
\begin{sketch} The theorem is proved by applying the following lemma, which is Theorem 9 in \cite{Prekopa80:Logarithmic}.

\begin{lem}\label{thm9}
Suppose $g_1(\vec{\theta},\vec{\zeta}),...,g_R(\vec{\theta},\vec{\zeta})$ are concave functions in $\mathbb{R}^{2m}$ where $\vec{\theta}$ is the vector of $m$ parameters and $\vec{\zeta}$ is a vector of $m$ real numbers that are generated according to a distribution whose pdf is logarithmic concave in $\mathbb{R}^m$. Then the following function is log-concave in $\mathbb{R}^m$.
 \begin{align}\label{joint}
 L_i(\vec{\theta},G)=\Pr(g_1(\vec{\theta},\vec{\zeta})\ge0,...,g_R(\vec{\theta},\vec{\zeta})\ge0), \ \ \vec{\theta}\in \mathbb{R}^m
 \end{align}
\end{lem}
To apply Lemma~\ref{thm9}, we define a set $G^i$ of function $g^i$'s that is equivalent to an order $\pi^i$ in the sense of inequalities implied by RUM for $\pi^i$ and $G^i$ (the joint probability in (\ref{joint}) for $G^i$ to be the same as the probity of $\pi^i$ in RUM with parameters $\vec{\theta}$).
Suppose $g^i_{r}(\vec{\theta},\vec{\zeta})=\theta_{\pi^i(r)}+\zeta^i_{\pi^i(r)}-\theta_{\pi^i{(r+1)}}-\zeta^i_{{\pi^i({r+1})}}$ for $r=1,..,m-1$.

Then considering that the length of order $\pi^i$ is $R+1$, we have:
\begin{align}
L_i(\vec{\theta},\pi^i)=L_i(\vec{\theta},G^i)=\Pr(g^i_1(\vec{\theta},\vec{\zeta})\ge0,...,g^i_R(\vec{\theta},\vec{\zeta})\ge0), \ \ \vec{\theta}\in \mathbb{R}^m
\end{align}

This is because $g^i_{r}(\vec{\theta},\vec{\zeta})\ge0$ is equivalent to that in $\pi^i$ alternative $\pi^i(r)$ is preferred to alternative $\pi^i(r+1)$ in the RUM sense.

To see how this extends to the case where preferences are specified
as partial orders, we consider in particular an interpretation where
an agent's report for the ranking of $m_i$ alternatives implies that
all other alternatives are worse for the agent, in some undefined
order. Given this, define
$g^i_{r}(\vec{\theta},\vec{\zeta})=\theta_{\pi^i(r)}+\zeta^i_{\pi^i(r)}-\theta_{\pi^i{(r+1)}}-\zeta^i_{{\pi^i({r+1})}}$
for $r=1,..,m_i-1$ and
$g^i_{r}(\vec{\theta},\vec{\zeta})=\theta_{\pi^i(m_i)}+\zeta^i_{\pi^i(m_i)}-\theta_{\pi^i{(r+1)}}-\zeta^i_{{\pi^i({r+1})}}$
for $r=m_i,..,m-1$.
Considering that $g^i_r(\cdot)$s are linear (hence, concave) and using log concavity of the distributions of $\vec {\zeta^i}=(\zeta^i_1,\zeta^i_2,..,\zeta^i_m)$'s, we can apply Lemma~\ref{thm9} and prove log-concavity of the likelihood function.
\end{sketch}

It is not hard to verify that pdfs for normal and Gumbel
are log-concave  under reasonable conditions for their parameters, made
explicit in the following corollary.

\begin{crol} For the location family where each $\zeta_j$ is a normal distribution with mean zero and with fixed variance, or Gumbel distribution with mean zeros and fixed shape parameter, $l(\vec{\theta};D)$ is concave. Specifically, the log-likelihood function for P-L is concave.
\end{crol}
The concavity of log-likelihood of P-L has been proved~\cite{Ford57:Solution} using a different technique. Using Fact 3.5.~in~\cite{Proschan89:Log}, the set of global maxima solutions to the likelihood function, denoted by $S_D$, is convex since the likelihood function is log-concave. However, we also need that $S_D$ is bounded, and would further like that it provides one unique order as the estimation for the ground truth.

For P-L, Ford, Jr.~\cite{Ford57:Solution} proposed the following necessary and sufficient condition for the set of global maxima solutions to be bounded (more precisely, unique) when $\sum_{j=1}^m e^{\theta_j}=1$.

\begin{cond}\label{cond1} Given the data $D$,
in every partition of the alternatives $\mc$ into two nonempty subsets $\mc_1\cup \mc_2$,  there exists
$c_1\in \mc_1$ and $c_2\in \mc_2$ such that there
is at least one ranking in $D$ where $c_1\succ c_2$.

\end{cond}

We next show that Condition~\ref{cond1} is also a necessary and sufficient condition for the set of global maxima solutions $S_D$ to be bounded in location families, when we set one of the values $\theta_j$ to be $0$ (w.l.o.g.,~let $\theta_1=0$). If we do not bound any parameter, then $S_D$ is unbounded, because for any $\vec\theta$, any $D$, and any number $s\in \mathbb R$, $l(\vec \theta;D)=l(\vec \theta+s;D)$.
\begin{thm}\label{thm2}
Suppose we fix $\theta_1=0$. Then, the set $S_D$ of global maxima solutions to $l(\theta;D)$ is bounded if and only if the data $D$ satisfies Condition~\ref{cond1}.
\end{thm}
\begin{sketch}

If Condition~\ref{cond1} does not hold, then $S_D$ is unbounded because the parameters for all alternatives in $C_1$ can be increased simultaneously to improve the log-likelihood. For sufficiency, we first present the following lemma.
\vspace{-2mm}
\begin{lem}\label{lem1}
If alternative $j$ is preferred to alternative $j'$ in at least in one ranking then the difference of their mean parameters $\theta_{j'}-\theta_j$ is bounded from above ($\exists Q \ where \ \theta_{j'}-\theta_j<Q$) for all the $\vec{\theta}$ that maximize the likelihood function.
\end{lem}
\vspace{-2mm}
\begin{proof}
Suppose that $j\succ j'$ in rank $i$, then for any $\vec{\theta}\in \mathbb{R}^m$:
\begin{align}
&L_i(\vec{\theta},\pi^i)=L_i(\vec{\theta},G^i)=\Pr(g_1(\vec{\theta},\vec{\zeta})\ge0,...,g_R(\vec{\theta},\vec{\zeta})\ge0)\nonumber\\
 \le & \Pr(g_{\pi^i(r)}(\vec{\theta},\vec{\zeta})\ge0, g_{\pi^i(r+1)}(\vec{\theta},\vec{\zeta})\ge0, \ldots, g_{\pi^i(r')}(\vec{\theta},\vec{\zeta})\ge0) \le
\Pr( \zeta_{j}-\zeta_{j'}\ge \theta_{j'}-\theta_{j}),
\end{align}
where $j=\pi^i(r)$ and $j'=\pi^i(r')$.

Let $K=l(\vec 0; D)$. Since the log-likelihood is always smaller than $0$, it follows that for any $\vec \theta\in S_D$ and any $i\leq n$, $L_i(\vec \theta;\pi^i)\geq K$.

Hence, $\Pr(\zeta_{j}-\zeta_{j'}\ge \theta_{j'}-\theta_{j})\geq K$.

Therefore, there exists $K'$ such that $\theta_{j'}-\theta_{j}< K'$, where $K'$ depends on the fixed $\zeta_{j'}$ and $\zeta_{j}$.
\end{proof}

Now consider a directed graph $G_D$, where the nodes are the
alternatives, and there is an edge between $c_{j}$ to $c_{j'}$ if in
at least one ranking $c_{j}\succ c_{j'}$.  By Condition~\ref{cond1},
for any pair $j\neq j'$, there is a path from $c_{j}$ to $c_{j'}$ (and
conversely, a path from $c_{j'}$ to $c_j$).  To see this, consider
building a path between $j$ and $j'$ by starting from a partition with
${\mathcal C}_1= \{j\}$ and following an edge from $j$ to $j_1$ in the
graph where $j_1$ is an alternatives in ${\mathcal C}_2$ for which
there must be such an edge, by Condition~\ref{cond1}. Consider the
partition with ${\mathcal C}_1=\{j,j_1\}$, and repeat until an edge
can be followed to vertex $j'\in {\mathcal C}_2$.
It follows from Lemma~\ref{lem1} that for any $\vec{\theta}\in S_D$  we have $|\theta_j-\theta_{j'}|<Qm$, using the telescopic sum of bounded values of the difference of mean parameters along the edges of the path, since the length of the path is no more than $m$ (and tracing the path from $j$ to $j'$ and $j'$ to $j$),
meaning that $S_D$ is bounded.
\end{sketch}

Now that we have the log concavity and bounded property, we need to declare conditions under which the bounded convex space of estimated parameters corresponds to a unique order. The next theorem provides a necessary and sufficient condition for all global maxima to correspond to the same order on alternatives.
Suppose that we order the alternatives based on estimated $\theta$'s (meaning that $c_j$ is ranked higher than $c_{j'}$ iff $\theta_j>\theta_{j'}$).

\begin{thm}\label{thm:sameorder}
The order over parameters 
is strict
and is the same across all $\vec\theta\in S_D$ if, for all $\vec\theta\in S_D$ and all alternatives
$j\neq j'$, $\theta_j\neq \theta_{j'}$.
\end{thm}
\vspace{-2mm}
\begin{proof}
Suppose for the sake of contradiction there exist two maxima, $\vec \theta, \vec\theta^*\in S_D$ and a pair of alternatives $j\neq j'$  such that $\theta_j>\theta_{j'}$ and $\theta^*_{j'}>\theta_{j}^*$. Then, there exists an $\alpha<1$ such that the $j$th and $j'$th components of $\alpha\vec \theta+(1-\alpha)\vec\theta^*$ are equal, which contradicts the assumption.
\end{proof}

Hence, if there is never a tie in the scores in any $\vec \theta\in S_D$, then any vector in $S_D$ will reveal the unique order.

\section{Monte Carlo EM for Parameter Estimation}

In this section, we propose an MC-EM algorithm for MLE inference for RUMs where every $\mu_j$ belongs to the EF.\footnote{Our algorithm can be naturally extended to compute a maximum {\em a posteriori} probability (MAP) estimate, when we have a prior over the parameters $\vec{\theta}$. Still, it seems hard to motivate the imposition of a prior on parameters in many social choice domains.}

The EM algorithm determines the MLE parameters $\vec{\theta}$ iteratively, and proceeds as follows. In each iteration $t+1$, given parameters $\vec{\theta}^t$ from the previous iteration, the algorithm is composed of an E-step and an M-step.  For the E-step, for any given $\vec\theta=(\theta_1,\ldots,\theta_m)$, we compute the conditional expectation of the complete-data log-likelihood (latent variables $\vec x$ and data $D$), where the latent variables $\vec x$ are distributed according to data $D$ and parameters $\vec{\theta}^t$ from the last iteration.

For the M-step, we optimize $\vec \theta$ to maximize the expected log-likelihood computed in the E-step, and use it as the input $\vec\theta^{t+1}$ for the next iteration:
\vspace{-3mm}
\begin{align}
&\mbox{E-Step :} \notag \ \ \ Q(\vec{\theta},\vec{\theta}^t)=E_{\vec X}\left\{\log \prod_{i=1}^n \Pr(\vec{x}^i,\pi^i \ |\ \vec{\theta})\ |\ D, \vec{\theta}^{t}\right\}
\\
&\mbox{M-step :}\notag \ \ \ \vec{\theta}^{t+1}\in\arg\max_{\vec{\theta}}
Q(\vec{\theta},\vec{\theta}^t)
\end{align}
\vspace{-5mm}
\subsection{Monte Carlo E-step by Gibbs sampler}
The E-step can be simplified using~\eqref{expf} as follows:
\begin{align*}
&E_{\vec X}\{\log \prod_{i=1}^n \Pr(\vec{x}^i,\pi^i\ |\ \vec{\theta})\ |\ {D, \vec{\theta}^{t}} \}=E_{\vec X}\{\log \prod_{i=1}^n \Pr(\vec{x}^i|\ \vec{\theta})\Pr(\pi^i|\vec{x}^i)\ |\ {D, \vec{\theta}^{t}} \}
\end{align*}
\begin{align*}
&=\sum_{i=1}^n \sum_{j=1}^m E_{X^i_j}\{\log \mu_j(x^i_j|\theta_j)\ |\  \pi^i,\vec{\theta}^{t}\}
=\sum_{i=1}^n \sum_{j=1}^m (\eta(\theta_j)E_{X^i_j}\{T(x^i_j)\ |\ \pi^i,\vec{\theta}^{t}\}-A(\theta_j)+W,
\end{align*}
where $W=E_{X^i_j}\{B(x^i_j)\ |\ \pi^i,\vec{\theta}^{t} \}$ only depends on $\vec\theta_t$ and $D$ (not on $\vec \theta$), which means that it can be treated as a constant in the M-step. Hence, in the E-step we only need to compute $S^{i,t+1}_j=E_{X_j^i}\{T(x^i_j)\ |\ \pi^i,\vec{\theta}^t\}$ where $T(x^i_j)$ is the sufficient statistic for the parameter $\theta_j$ in the model. We are not aware of an analytical solution for $E_{X^i_j}\{T(x^i_j)\ |\ \pi^i,\vec{\theta}^t\}$. However, we can use a Monte Carlo approximation, which involves sampling $\vec{x}^i$ from the distribution $\Pr(\vec{x}^i\ |\ \pi^i,\vec{\theta^t})$ using a Gibbs sampler, and then approximates $S^{i,t+1}_j$ by ${1\over N}\sum^N_{k=1}T(x^{i,k}_j)$ where $N$ is the number of samples in the Gibbs sampler.

\begin{figure}[h!]
\begin{minipage}[c]{0.5\linewidth}

\centering
  \includegraphics[width=.9\textwidth]{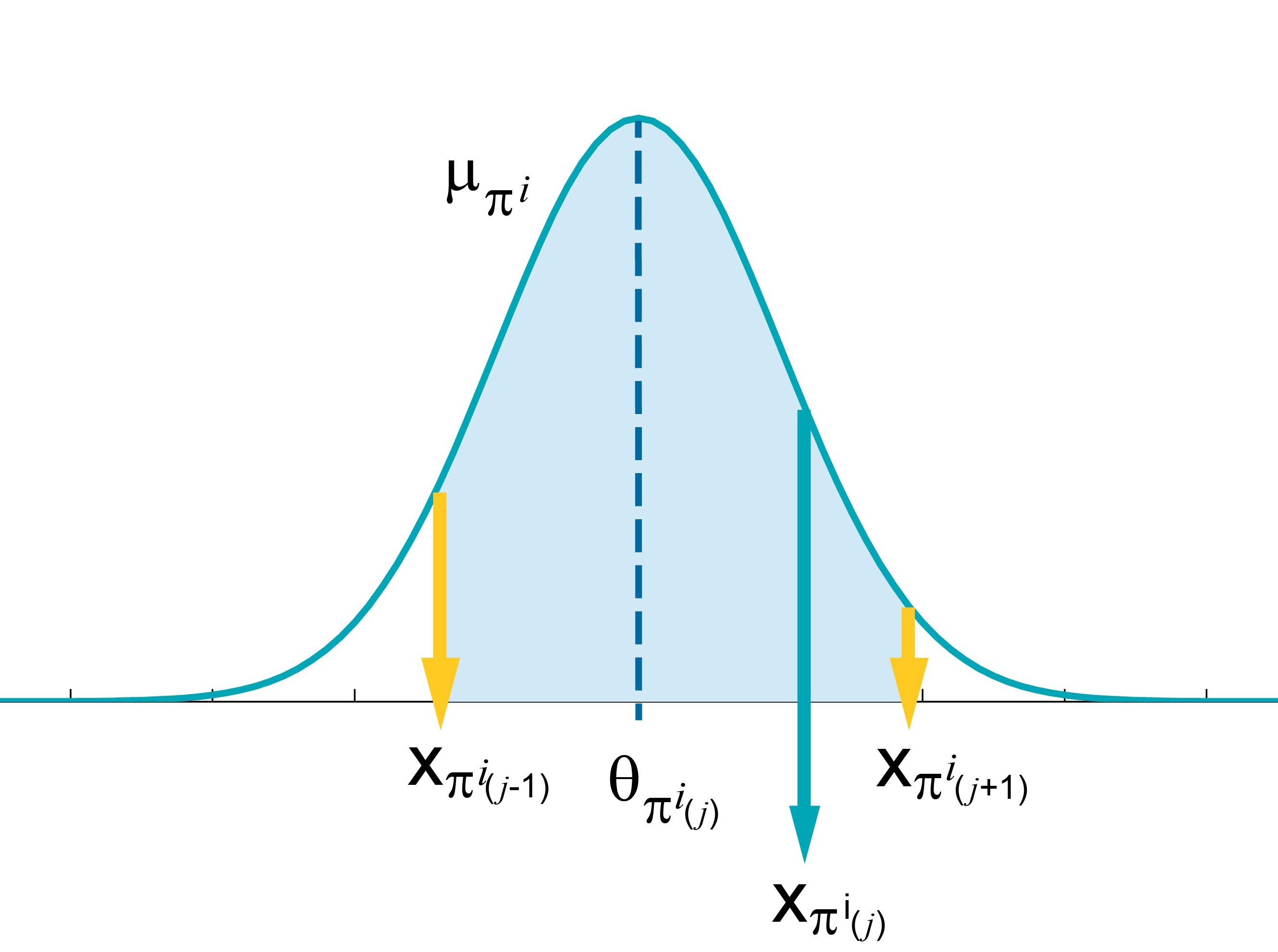}\vspace{-4mm}

 \caption{\small A truncated normal distribution. \label{figreTN}}
\end{minipage}
\begin{minipage}[c]{0.5\linewidth}
\textbf{Rao-Blackwellized}: To further improve the Gibbs sampler, we use Rao-Blackwellized \cite{Brooks11:Handbook} estimation using $E\{T(x^{i,k}_j)\ |\ x^{i,k}_{-j},\pi^i,\vec{\theta}^t\}$ instead of the sample $x^{i,k}_j$, where $x^{i,k}_{-j}$ is all of $\vec{x}^{i,k}$ except for $x^{i,k}_j$.
Finally,  we estimate $E\{T(x^{i,k}_j)\ |\ x^{i,k}_{-j},\pi^i,\vec{\theta}^t\}$ in each step of the Gibbs sampler using $M$ samples as
$S_j^{i,t+1}\simeq{1\over N}\sum^N_{k=1} E\{T(x^{i,k}_j)\ |\ x^{k}_{-j},\pi^i,\vec{\theta}^t\}\simeq{1\over NM}\sum^N_{k=1} \sum^M_{l=1} T(x^{i^l,k}_j),$
where $x^{i^l,k}_j\sim \Pr(x^{i^l,k}_j\ |\ x^{i,k}_{-j}, \pi^i,\vec{\theta})$.  Rao-Blackwellization reduces the variance of the estimator because of conditioning and expectation in $E\{T(x^{i,k}_j)\ |\ x^{i,k}_{-j},\pi^i,\vec{\theta}^t\}$.
\end{minipage}
\end{figure}

In each step of our Gibbs sampler for voter $i$, we randomly choose a position $j$ in $\pi^i$ and sample $x_{\pi^i(j)}^i$ according to a {\em TruncatedEF} distribution $\Pr(\cdot |\ x_{\pi^i(-j)}, \vec{\theta^t}, \pi^i)$, where $\ x_{\pi^i(-j)}=(\ x_{\pi^i(1)},\ldots, \ x_{\pi^i(j-1)}, \ x_{\pi^i(j+1)},\ldots,\ x_{\pi^i(m)})$.
The TruncatedEF is obtained by truncating the tails of $\mu_{\pi^i(j)}(\cdot|\theta^t_{\pi^i(j)})$ at $x_{\pi^i(j-1)}$ and $x_{\pi^i(j+1)}$, respectively. For example, a truncated normal distribution is illustrated in Figure~\ref{figreTN}.
%

\vspace{-6mm}
\subsection{M-step}
In the E-step we have (approximately) computed $S_j^{i,t+1}$. In the M-step  we compute $\vec\theta^{t+1}$ to maximize $\sum_{i=1}^n \sum_{j=1}^m (\eta(\theta_j)E_{X^i_j}\{T(x^i_j)\ |\ \pi^i,\vec{\theta}^{t}\}-A(\theta_j)+E_{X^i_j}\{B(x^i_j)\ |\ \pi^i,\vec{\theta}^{t}\})$.
Equivalently, we compute $\theta_j^{t+1}$ for each $j\leq m$ separately to maximize $\sum_{i=1}^n\{\eta(\theta_j)E_{X^i_j}\{T(x^i_j)\ |\ \pi^i,\vec{\theta}^{t}\}-A(\theta_j)\}=\eta(\theta_j)\sum_{i=1}^nS_{j}^{i,t+1}-nA(\theta_j)$.
 
\begin{figure*}[h]
  \centering
\includegraphics[width=400pt]{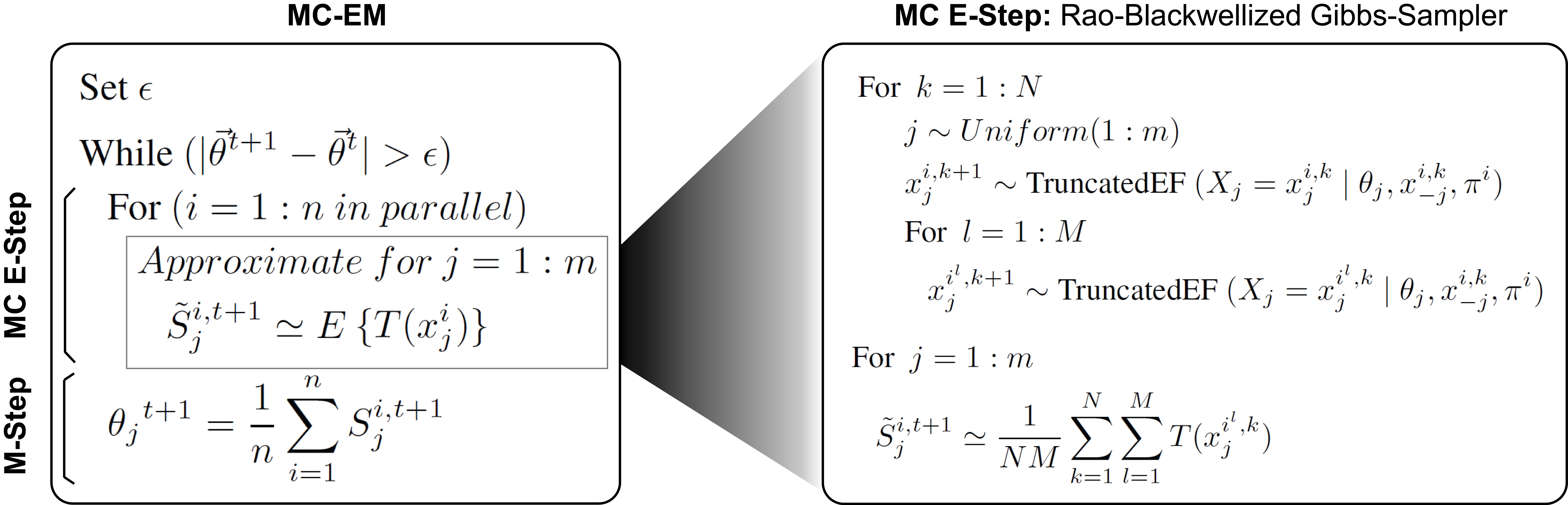}\vspace{-2mm}
  \caption{\small The MC-EM algorithm for normal distribution.\label{figalg}}
\end{figure*}

For the case of the normal distribution with fixed variance, where $\eta(\theta_j)=2 \theta_j$ and $A(\theta_j)=(\theta_j)^2$, we have $\theta_{j}^{t+1}={1\over n}\sum^n_{i=1} S^{i,t+1}_j$. The algorithm is illustrated in Figure~\ref{figalg}.
%
\subsection{Convergence}

In the last section we showed that if the RUM satisfies the premise in
Theorem~\ref{thm1} and Theorem~\ref{thm2}
the data satisfies Condition~\ref{cond1}, then the log-likelihood function is concave, and the set of global maxima solutions is bounded. This guarantee the convergence of MC-EM for an exact E-step.

In general, MC-EM methods do not have the uniform convergence property of EM methods. In order to control the error of approximation in the MC-E step we can increase the number of samples with the iterations~\cite{Wei90:Monte}. However, in our application, we are not concerned with the exact estimation of $\vec{\theta}$, as we are only interested in their orders relative to each-other. Therefore, as long as the approximation error remains relatively small, such that the differences of $\theta_j$s are much larger than the error, we are safe to stop.

A known problem with Gibbs sampling is that it can introduce correlation among samples. To address this, we sub-sample the samples to reduce the correlation, and call the ratio of sub-sampling the {\em thinning factor} ($0< F\le1$). A suitable thinning ratio can be set using empirical results from the sampler.

With an approach similar to \cite{Booth99:Maximizing}, we can derive a relationship between the variance of error in $\vec{\theta}^{t+1}$ and the Monte-Carlo error in the E-step approximation:
\begin{align}
\mathit{Var}({\theta_j}^{t+1})={1\over n^2}\sum^n_{i=1} \mathit{Var}(S^{i,t+1}_j)={1\over MNn^2}\sum^n_{i=1} \mathit{Var}(x^{i}_j)\le {F V\over MNn},
\end{align}
where $N$ is number of samples in Gibbs sampler, $M$ is the number of samples for Rao-Blackwellization, $n$ is number of agents, $F$ is the thinning factor and $V=\max_j (\mathit{Var}_{x \sim \mu_j}(x))$, and samples $x^{i}_j$ are assumed to be independent. Given, $T$, $V$ and $n$, we can make $\mathit{Var}({\theta_j}^{t+1})$ arbitrarily small by increasing $MN$.

\vspace{-2mm}
\section{Experimental Results}\vspace{-2mm}

We evaluate the proposed MC-EM algorithm on synthetic data as well as two real world data sets, namely an election data set and a dataset representing preference orders on sushi. For simulated data we use the Kendall correlation~\cite{Grzegorzewski09:Kendall} between two rank orders (typically between the true order
and the method's result) as a measure of performance.
\vspace{-2mm}
\subsection{Experiments for Synthetic Data}\vspace{-2mm}

\begin{figure}[htp]

\includegraphics[width=1\textwidth]{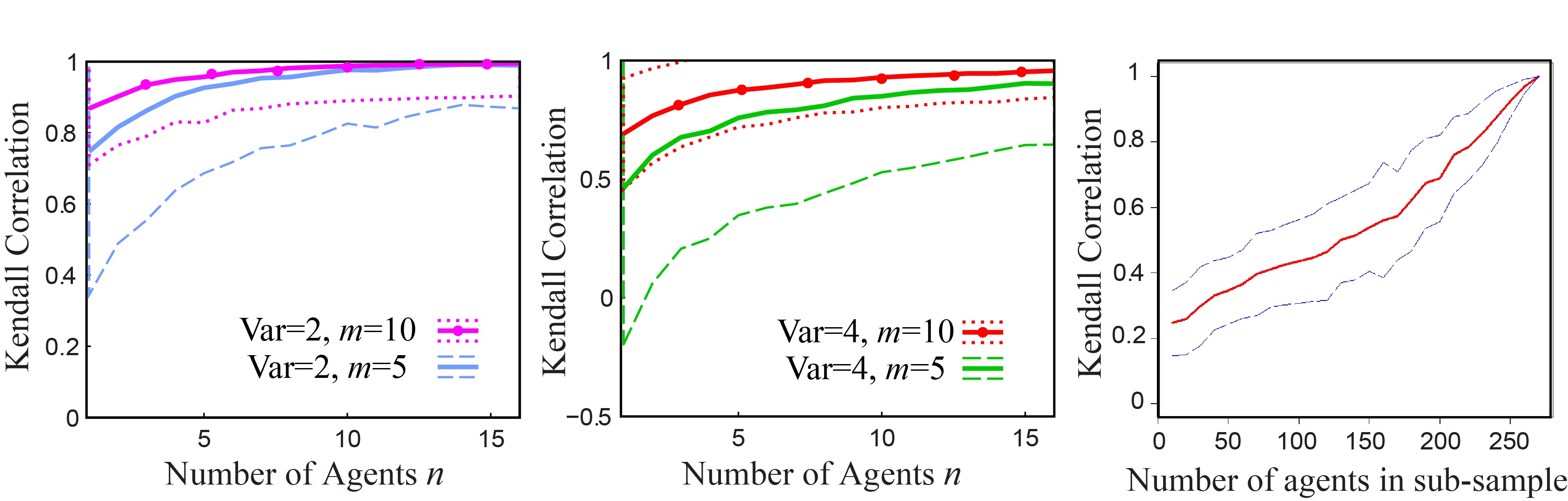}

\begin{minipage}[l]{1\linewidth}
\caption{\small Left and middle panel: Performance for different number of agents $n$
on synthetic data for $m=5,10$  and $\mathit{Var}=2,4$, with specifications $MN=4000$, $EM iterations=3$. Right panel: Performance given access to sub-samples of the data in the public election dataset,  x-axis: size of sub-samples, y-axis: Kendall Correlation with the order obtained from the full data-set. Dashed lines are the 95\% confidence intervals. \label{figcomp}}
\end{minipage}
\end{figure}

We first generate data from Normal models for the random utility terms, with means $\theta_j=j$ and equal variance for all terms, for different choices of variance ($\mathit{Var}=2,4$). We evaluate the performance of the method as the number of  agents $n$ varies. The results show that a limited number of iterations in the EM algorithm (at most 3), and samples $MN=4000$ (M=5, N=800) are sufficient for inferring the order in most cases. The performance in terms of Kendall correlation for recovering ground truth improves for larger number of agents, which corresponds to more data. See Figure~\ref{figcomp}, which shows the asymptotic behavior of the maximum likelihood estimator in recovering the true parameters. Figure~\ref{figcomp} left and middle panels show that the more the size of dataset the better the performance of the method.

Moreover, for large variances in data generation, due to increasing noise in the data, the rate that performance gets better is slower than that for the case for smaller variances. Notice that the scales on the y-axis are different in the left and middle panels.

\vspace{-2mm}

\subsection{Experiments for Model Robustness}

We apply our method to  a public election dataset collected by Nicolaus Tideman~\cite{Tideman06:Collective}, where the voters provided partial orders on candidates. A partial order includes comparisons among a subset of alternative, and the non-mentioned alternatives in the partial order are considered to be ranked lower than the lowest ranked alternative among mentioned alternatives.

The total number of votes are $n=280$ and the number of alternatives $m=15$.
For the purpose of our experiments, we adopt the order on alternatives obtained by applying our method on the entire dataset  as an assumed ground truth, since no ground truth is given as part of the data. After finding the ground truth by using all 280 votes (and adopting a normal model), we compare the performance of our approach as we vary the amount of data available. We evaluate the performance for sub-samples consisting of $10,20,\ldots,280$ of samples randomly chosen from the full dataset. For each sub-sample size, the experiment is repeated $200$ times and we report the average performance and the variance. See the right panel in Figure~\ref{figcomp}.
This experiment shows the robustness of the method,
 in the sense that the result of inference on a subset of the dataset shows consistent behavior with the case that the result on the full dataset. For example, the ranking obtained by using half of the data can still achieve a fair estimate to the results with full data, with an average Kendall correlation of greater than 0.4.

\subsection{Experiments for Model Fitness}

In addition to a public election dataset, we have tested our
algorithm on a sushi dataset, where 5000 users give rankings over 10
different kinds of sushi~\cite{Kamishima03:Nantonac}. For each
experiment  we randomly choose $n\in\{10, 20, 30, 40,
50\}$ rankings, apply our MC-EM for RUMs with normal distributions
where variances are also parameters.

In the former experiments, both the synthetic data generation and the model for election data,
the variances were fixed to $1$ and hence we had the theoretical guarantees for the convergence to global optimal solutions by Theorem~\ref{thm1} and Theorem~\ref{thm2}. When we let the variances to be part of parametrization we lose the theoretical guarantees. However, the EM algorithm can still be applied, and since the variances are now parameters (rather than being fixed to $1$), the model fits better in terms of log-likelihood.

For this reason, we adopt RUMs with normal distributions in which the variance is a parameter
that is fit by EM along with the mean. We call this model a {\em normal model}.
%
We compute the difference between the normal model and P-L in terms of four criteria: log-likelihood (LL), predictive log-likelihood (predictive LL), AIC, and BIC. For (predictive) log-likelihood, a positive value means that normal model fits better than P-L, whereas for AIC and BIC, a negative number means that normal model fits better than P-L.
Predictive likelihood is different from likelihood in the sense that we compute the likelihood of the estimated parameters for a part of the data that is not used for parameter estimation.\footnote{The use of predictive likelihood allows us to evaluate the performance of the estimated parameters on the rest of the data, and is similar in this sense to cross validation for supervised learning.} In particular, we compute predictive likelihood for a randomly chosen subset of $100$ votes. The results and standard deviations for $n=10,50$ are summarized in Table~\ref{tab:difference}.\vspace{-2mm}

\begin{table}[htp]
\centering
\begin{tabular}{|c|@{\hspace{1pt}}c@{\hspace{1pt}}|@{\hspace{1pt}}c@{\hspace{1pt}}|@{\hspace{1pt}}c@{\hspace{1pt}}|@{\hspace{1pt}}c@{\hspace{1pt}}|c@{\hspace{1pt}}|@{\hspace{1pt}}c@{\hspace{1pt}}|@{\hspace{1pt}}c@{\hspace{1pt}}|@{\hspace{1pt}}c@{\hspace{1pt}}|c|}
\hline
&\multicolumn{4}{c|}{$n=10$\hspace{-5cm}}&\multicolumn{4}{c|}{$n=50$\hspace{-2cm}}\\
\hline
Dataset&\small LL&\small Pred.~LL&\small AIC&\small BIC&\small LL&\small Pred.~LL&\small AIC&\small BIC\\
\hline
Sushi& \small \bf 8.8(4.2)&\small -56.1(89.5) &\small-7.6(8.4) &\small  5.4(8.4)&\small  \bf  22.6(6.3)&\small  \bf  40.1(5.1)&\small \bf  -35.2(12.6)&\small -6.1(12.6)\\
\hline
Election& \small 9.4(10.6)&\small 91.3(103.8)&\small -8.8(21.2) &\small 4.2(21.2) &\small \bf  44.8(15.8)&\small \bf  87.4(30.5)&\small \bf -79.6(31.6)&\small -50.5(31.6) \\
\hline
\end{tabular}
\caption{\small Model selection for the sushi dataset and election dataset. Cases where the normal model fits better than P-L statistically with 95\% confidence are in bold.\label{tab:difference}}
\end{table}

When $n$ is small $(n=10)$, the variance is high and we are unable to obtain statistically significant results in comparing fitness.
When $n$ is not too small ($n=50$), RUMs with normal distributions fit better than P-L.
Specifically, for log-likelihood, predictive log-likelihood, and AIC, RUMs with normal distributions outperform P-L with 95\% confidence in both datasets.

\subsection{Implementation and Run Time}

The running time for our MC-EM algorithm scales linearly with number of agents on real world data (Election Data)
with slope 13.3 second per agent on an Intel $i5$ $2.70$GHz PC. This is for 100 iterations of EM algorithm with Gibbs sampling number increasing with iterations as $2000+300*iteration \ steps$.
%
%

\vspace{-3mm}
\section*{Acknowledgments}
\vspace{-3mm}

This work is supported in part by NSF Grant No. CCF- 0915016. Lirong Xia is supported by NSF under Grant \#1136996 to the Computing Research Association for the CIFellows Project. We thank Craig Boutilier, Jonathan Huang, Tyler Lu, Nicolaus Tideman, Paolo Viappiani, and anonymous NIPS-12 reviewers for helpful comments and suggestions, or help on the datasets.

{\small

}

\end{document}